\documentclass[10pt,twocolumn, prl, showpacs]{revtex4-1}
\usepackage{graphicx}
\usepackage{amsmath}
\usepackage{latexsym,amssymb,epsfig,graphicx}
\begin{document}
\title{Macroscopic view of light pressure on continuous medium}
\author{ M.V. Gorkunov and A.V. Kondratov}
\affiliation{A.V. Shubnikov Institute of Crystallography, Russian Academy of Sciences, 119333 Moscow, Russia}

\begin{abstract}
The ambiguity of macroscopic description of light pressure on continuous medium originates from the uncertainty of dividing the energy-momentum tensor of electromagnetically excited matter into material and field parts or, equivalently, the total acting force into pressure and deformation terms. We show that although there exists a continuum of formally correct formulations, one can adopt the appropriate form of the macroscopic field stress tensor that allows unified description of pressure during elementary light-matter interactions, such as  reflection/refraction, absorption and nonlinear conversion. The proposed expressions for the pressure force are simple, convenient and compatible with the polariton momentum $\hbar \bf k$. The corresponding electromagnetic momentum density \eqref{G} generalizes Rytov's definition for right-handed and left-handed frequency dispersive media.
\end{abstract}
\pacs{42.50.Wk; 42.25.Bs; 78.20.Bh}
\maketitle

Since the pioneering works by Lebedev \cite{Lebedev}, the light pressure has been the most substantial observable consequence of the momentum transfer by electromagnetic (EM) waves in vacuum. For decades afterwards, the phenomenon was supposed to be primarily of general importance with uncertain practical value reduced to e.g. gigantic solar-sail propelled spaceships \cite{Clarke}.
Later it became clear that observable and useful manifestations of the light pressure occur actually on the submillimeter scale, where powerful laser beams can be  used as tweezers for microparticles \cite{Ashkin1970}. Nowadays, the optomechanics has developed into a broad area with numerous potential applications in nanotechnology (see e.g. \cite{Marquardt}).

In the same way as for 'pure' photons in vacuum, the concept of light pressure in continuous medium has to rely on momentum carried by the corresponding quasiparticles -- polaritons. However, the definition of the polariton momentum remains the subject of ongoing discussions. The crucial obstacle is the lack of established unique definitions of momentum density and flux of EM-field in continuous medium.

Over the years, various definitions have been proposed and analyzed \cite{Abraham,Minkovski,Rytov,
Ginzburg,Mansuripur1,BarnettPRL,Mansuripur2,Kemp,Pfeifer,Shevchenko2,Veselago1968}. Two simplest and most prominent variants belong to Minkovski and Abraham. According to  Abraham \cite{Abraham}, the momentum density of EM-wave reads ${\bf G}_A={\bf
E\times H}/(4\pi c)$, while Minkovski's alternative is ${\bf G}_M={\bf D\times B}/(4\pi c)$ \cite{Minkovski}. The latter has been generalized by Rytov \cite{Rytov} for substantial frequency dispersion involving terms with frequency derivatives of permeability and permittivity (see also \cite{Kemp,Veselago1968}). More recently, other less trivial variants have been proposed (e.g., a half-sum of ${\bf G}_A$ and ${\bf G}_M$ \cite{Mansuripur1}) although it has been argued that only ${\bf G}_A$ and ${\bf G}_M$  possess special physical meanings and other 'rival forms' are excessive \cite{BarnettPRL}.

Surprisingly, no direct controversy seems to follow from any reasonable definition, which supports the common belief that {``\it the preferred form is therefore effectively a matter of personal choice''}\cite{Pfeifer}. To avoid this uncertainty when solving practical problems (e.g. calculating the forces in confined fields \cite{Novotny}) one often encloses a finite object (microparticle) by a surface in vacuum and uses the vacuum relations to obtain the overall force and torque. Then, however, the details of the momentum transfer remain unrevealed and one cannot trace the contributions arising from reflection/refraction at the surfaces, absorption in the volume or finite pulse duration or explain the roles of key parameters (particle shape, refractive index, absorption, etc.).

In some applied domains the above unclarity has not even been fully recognized. Thus throughout decades in the literature on
nonlinear optics the phase matching condition is mixed with the light momentum conservation law (see
e.g. the classical \cite{Armstrong,Landau}, or the
recent \cite{Bahabad}), which
imposes the polariton momentum directly expressed by the polariton wave vector as $\hbar {\bf k}$.
However, the link between phase matching and momentum transfer during nonlinear processes has not been established.

The topic has got recently another intriguing dimension with emergence of metamaterials. While in conventional dielectrics the ambiguity of polariton momentum can be seen as a minor quantitative issue, in the left-handed media (LHM) with negative permeability and permittivity, the vectors of polariton phase and group velocities are antiparallel \cite{Veselago1968, Pendry} and it becomes a matter of momentum and pressure signs \cite{Yannopapas, Kemp, Veselago2009}.

In this Letter our aims are: (i) to analyze the actual degree of freedom in defining light momentum and pressure in continuous medium and to show that there exists a {\it continuum} of possibilities of formally self-consistent description fulfilling the necessary conservation laws; (ii) to propose a unified approach that yields simple and convenient expressions for the light pressure during elementary light-matter interactions. In this approach, the forces acting on inhomogeneous medium, as well as during light reflection/refraction and absorption, pulse propagation and nonlinear conversion can all be presented by simple self-consistent expressions that are compatible with the polariton momentum $\hbar {\bf k}$ and the EM-field momentum density as simple as Eq.~\eqref{G} applicable to both right-handed and left-handed frequency dispersive media.

Several relations on the microscopic level form the basis for the following macroscopic analysis. Thus
the microscopic forces acting on matter can be conveniently derived from the Maxwell's stress tensor \cite{Jackson}:
\begin{equation}\label{Tmicro}
{t}_{ij}=(4\pi)^{-1}\left[{e}_i{e}_j+{b}_i{b}_j-{\delta_{ij}}\left(e^2+b^2\right)/2\right],
\end{equation}
expressed by the microscopic electric and magnetic fields $\bf e({\bf r},t)$ and $\bf b({\bf r},t)$.
The stress tensor equates with the momentum flux tensor taken with the opposite sign, which can be illustrated by considering \eqref{Tmicro} for a monochromatic
plane wave in vacuum (wave vector $k=\omega/c$, fields  ${\bf e}=-c/\omega\ {\bf k\times b}$ and
${\bf b}=c/\omega\ {\bf k\times e}$):
\begin{equation}\label{Tmicrophys}
{t}^{\rm pw}_{ij}=(4\pi)^{-1}({\bf e\times b})_i{k_j}/{k}\equiv -g_iv_j.
\end{equation}
Here ${\bf v}=c{\bf k}/k$ is the group velocity  and ${\bf g}={\bf e}\times{\bf b}/(4 \pi c)$ is the microscopic density of momentum. The latter form presents the stress tensor as a dyadic product of the momentum density with the velocity of its transmission, i.e. as the momentum flux tensor.

To define the momentum quant in this otherwise classical picture one introduces the energy quant $\hbar\omega$ and
writes the energy flux and the momentum flux as
proportional to the photon number flux $\bf N$:  ${\bf S}={\bf e\times b}\ c/{4\pi}={\bf N}\hbar\omega$, and $t_{ij}=-{N}_i{p}_j$. Comparing with Eq.~\eqref{Tmicrophys}
yields the quant momentum ${\bf p}=\hbar{\bf k}$.

The divergence of Eq.\eqref{Tmicro} yields the microscopic momentum balance \cite{Jackson}:
\begin{equation}\label{balancemic}
\frac{\partial}{\partial x_j} t_{ij}={f}_{{\rm L}i}+\frac{\partial}{\partial t}g_i,
\end{equation}
where  ${\bf f}_{{\rm L}}$ is the microscopic Lorentz force density acting on microscopic charge and current densities $\rho_{\rm mic}$ and $\bf j_{\rm mic}$:
\begin{equation}\label{fL}
{\bf f}_L=\rho_{\rm mic}{\bf e}+({\bf j_{\rm
mic}}\times {\bf b})/c.
\end{equation}

Similarly, to derive the microscopic angular momentum density and torque one introduces the angular momentum tensor \cite{Jackson,Novotny,Landau2}:
\begin{equation}
m_{ij}=\delta_{ikn} x_k t_{nj},
\end{equation}
where $\delta_{ijk}$ is the antisymmetric Levi-Civita symbol. The microscopic angular momentum balance reads
\begin{equation}\label{balancemictorque}
\frac{\partial}{\partial x_j} m_{ij}=\phi_{{\rm L}i}+\frac{\partial}{\partial t}l_i,
\end{equation}
where  ${\boldsymbol  \phi}_{\rm L}={\bf r}\times {\bf f}_{{\rm L}}$ is the microscopic torque and $\bf l=r\times g$ is the angular momentum density.

For the EM-fields periodic in time, the last terms in Eqs.~(\ref{balancemic}, \ref{balancemictorque}) vanish upon time averaging and the average total force and torque acting on a finite body in vacuum can be evaluated from the stress tensor distribution outside the body as:
\begin{equation}\label{Ftot}
\bar{\bf F}_{\rm tot}=\int_V dV\ \bar{\bf f}_L = \int_S \bar{\bf t} \cdot d{\bf S},
\end{equation}
\begin{equation}\label{Phitot}
\bar{\boldsymbol \Phi}_{\rm tot}=\int_V dV \bar{\boldsymbol  \phi}_{\rm L} = \int_S \bar{\bf m} \cdot d{\bf S},
\end{equation}
respectively, with the integrals taken over the volume
$V$ including the body and the surface $S$ in vacuum enclosing the body, while the bar stands for the time averaging.

Macroscopic theory of light pressure has to comprise macroscopic analogues of Eqs. \eqref{balancemic} and \eqref{balancemictorque} with appropriate macroscopic force density. Unfortunately, a general macroscopic averaging of the above is impossible due to the quadratic dependence of the quantities on microscopic fields and densities of charge and current. Note that such an averaging can be performed within particular microscopic models of the medium (see e.g. \cite{Mansuripur2, Shevchenko1, Loudon}) but yields different model-specific results. This is not surprising since the procedure accounts for all forces arising in a body subjected to EM-fields: those responsible for the momentum exchange and those that tend to deform the body without momenta transfer.

The possibility of various formally self-consistent approaches is recognized to be the consequence of the ambiguity of separating the energy-momentum tensors of EM-field and matter \cite{Pfeifer,Ginzburg}. It is only the total tensor of the whole system that has a clear physical meaning. The separation line is uncertain but we find that only few general restrictions are sufficient to ensure the fulfillment of conservation laws. Namely, the coordinate part of the energy-momentum tensor -- the EM stress tensor (macroscopic analogue of Maxwell's tensor \eqref{Tmicro}) -- must: (i) be quadratic in the macroscopic fields $\bf E,B,D,H$; (ii) tend to Eq.~\eqref{Tmicro} in vacuum; (iii) be symmetric.

For any such a bilinear form ${\bf T}({\bf E,B,D,H})$ one can write the macroscopic momentum balance:
\begin{equation}\label{balancemac}
\frac{\partial}{\partial x_j} \bar{T}_{ij}=\bar{F}_{{\rm L}i},
\end{equation}
which serves then  as definition of the pressure force ${\bf F}_L$. For a finite body enclosed by a surface $S$,  ${\bf T}={\bf t}$ on $S$ (in vacuum), and the total force calculated as the integral $\int_V\bar{\bf F}_{\rm L}dV$ will always be equal to $\bar{\bf F}_{\rm tot}$ given by Eq.\eqref{Ftot} thus fulfilling the total momentum conservation automatically.

As usual in the field theory \cite{Landau2}, the conservation of the total angular momentum will also hold true provided that $T_{ij}$ is symmetric. Indeed, for the macroscopic angular momentum tensor $M_{ij}=\delta_{ikn} x_k T_{nj}$, one can introduce the macroscopic torque $\bar {\boldsymbol \Phi}_{\rm L}$ via:
\begin{equation}\label{balancemactorque}
\frac{\partial}{\partial x_j} \bar{M}_{ij}=\bar{\Phi}_{{\rm L}i},
\end{equation}
which equals the torque exerted by the pressure force, $\bar{\boldsymbol \Phi}={\bf r} \times{\bf \bar{F}}_{L}$, and sums up to the total torque $\int_V \bar{\boldsymbol \Phi}_{\rm L}dV$ that automatically equals Eq.~\eqref{Phitot} since  $\bar{\bf M}=\bar{\bf m}$ on the surface $S$ in vacuum.

Using the macroscopic force and torque defined from Eqs.~\eqref{balancemac} and \eqref{balancemactorque} one can consider elementary light-matter interactions, introduce an appropriate EM-wave momentum density to ensure the momentum conservation, etc. Obviously, there can be as many different formulations as there are possibilities to write the appropriate bilinear form ${\bf T}({\bf E,B,D,H})$, i.e.,
an {\it infinite continuum}. Existing theories as those in
Refs.~\cite{Kemp,Mansuripur1,Shevchenko2,BarnettTrans,Gordon} can all be obtained using this strategy. In this sense, they all are equally correct: however different spatial distributions of light pressure and torque they yield, the total force and torque acting on the whole body is the same.

For example, choosing the stress tensor $\left[{ E}_i {E}_j+{B}_i{B}_j-{\delta_{ij}}/2\
\left(E^2+B^2\right)\right]/{4\pi}$ yields the macroscopic pressure force $\rho{\bf E}+({\bf j}\times {\bf B})/c$, as if the macroscopic fields act on the macroscopic densities of bound charges $\rho$ and currents $\bf j$. Essentially, it is the similarity with \eqref{fL} that has  motivated the authors of Refs.~\cite{Gordon,Mansuripur1,BarnettTrans} to use such approach. However, it cannot really be obtained by macroscopic averaging of \eqref{fL} and its
severe drawbacks become obvious as one applies it to particular light-matter interactions. Thus the force is nonzero already for a superposition of EM-waves of the same frequency propagating through lossless homogeneous medium in different directions, i.e., the momentum exchange occurs in the absence of real interaction. Such 'virtual' terms must be then taken into account to ensure the momentum conservation and calculations become cumbersome.

A much clearer picture follows from the stress tensor :
 \begin{multline}\label{Tmacro}
{T}_{ij}=(8\pi)^{-1}[E_iD_j+D_iE_j+H_iB_j+B_iH_j-\\
\delta_{ij}\left({\bf E\cdot D}+{\bf
H\cdot B}\right)],
\end{multline}
first proposed by Rytov \cite{Rytov}.
Similar but unsymmetrized versions have been considered e.g. in Refs.~\onlinecite{Pfeifer, KempPRL} and for monochromatic fields in homogeneous media \eqref{Tmacro} reduces to the classical form \cite{Jackson, Novotny, Ginzburg}, but it is the expression \eqref{Tmacro} that yields correctly all the relations below.

Consider first Eq.~\eqref{Tmacro} for a monochromatic transversal plane wave traveling through an isotropic medium with $\varepsilon$ and
$\mu$ of the same sign, i.e., with the wavevector
$k=\sqrt{\varepsilon\mu}\ \omega/c$ and the fields ${\bf
D}=-c/\omega\ {\bf k\times H}$, ${\bf B}=c/\omega\ {\bf k\times
E}$, ${\bf D}=\varepsilon\bf E$, ${\bf B}=\mu\bf H$. The stress tensor \eqref{Tmacro} then reads
\begin{equation}\label{Tmacrophys}
{T}^{\rm pw}_{ij}=-\frac{c}{4\pi\omega}k_i({\bf E\times H})_j=-p_iN_j,
\end{equation}
where the polariton density
flux ${\bf N}={\bf
S}/(\hbar\omega)$ is expressed by the Poynting vector $\bf S$, and the latter form is the diadic representation
with the polariton momentum ${\bf p}=\hbar{\bf k}$.

On the other hand, the polariton dispersion law
$k^2=\varepsilon\mu\ \omega^2/c^2$ determines the group velocity
\begin{equation}\label{vgroup}
{\bf v}=\frac{\partial\omega({\bf k})}{\partial {\bf k}}= {\bf k}
\frac{2c^2}{\omega}
\left[\mu\frac{d(\omega\varepsilon)}{d\omega}+
\varepsilon\frac{d(\omega\mu)}{d\omega}\right]^{-1},
\end{equation}
and one can write \eqref{Tmacrophys} as ${T}^{\rm pw}_{ij}= - G_i v_j$ (compare with Eq.~\eqref{Tmicrophys})
with the momentum density of the EM wave
\begin{equation}\label{G}
{\bf G}=\frac{ 1}{8\pi c}({\bf E\times
H})\left[\mu\frac{d(\omega\varepsilon)}{d\omega}+
\varepsilon\frac{d(\omega\mu)}{d\omega}\right],
\end{equation}
This most general form reduces to Rytov's definition \cite{Rytov} in ordinary
media (for positive $\varepsilon$ and $\mu$) and further to  Minkovski's ${\bf G}_M$ upon neglecting the dispersion.

Next, taking space derivatives of \eqref{Tmacro} it is easy to obtain the macroscopic differential force-momentum balance with the total force
\begin{multline}\label{Ftot1}
{\bf F}=(8\pi)^{-1}[{\bf D({\nabla\cdot E})}+{\bf B({\nabla\cdot H})}-{\bf E}\times{\rm curl}{\bf  D}\\
-{\bf D}\times{\rm curl}{\bf  E}-{\bf H}\times{\rm curl}{\bf  B}-{\bf B}\times{\rm curl}{\bf  H}].
\end{multline}

For an arbitrary combination of monochromatic fields in a homogeneous isotropic transparent medium, this force reduces to the full time derivative ${\bf F}=(4\pi
c)^{-1}\ {\partial}/{\partial t}({\bf D\times B})$ and its time-average $\bar{\bf F}$ vanishes identically. Therefore, no virtual forces arise for a superposition of non-interacting EM-waves.

In a transparent and isotropic but inhomogeneous medium, \begin{equation}\label{Finhom}
\bar{\bf F}=-(8\pi)^{-1}\left(\overline{E^2}\nabla\varepsilon+\overline{H^2}\nabla\mu\right),
\end{equation}
i.e., the averaged force always points opposite the direction of the gradients of $\mu$ and $\varepsilon$, and, in particular, normally to interfaces.

To obtain the force acting on an abrupt $xy$-plane interface between two media, one can introduce a transient layer of infinitesimal thickness $\delta$ with $\varepsilon(z)$ and $\mu(z)$ continuously varying from $\varepsilon_1$ and $\mu_1$ at $z<-\delta/2$ to $\varepsilon_2$ and $\mu_2$ at $z>\delta/2$ respectively. Using the tangential field components ${\bf E}_\tau$ and ${\bf H}_\tau$ and normal induction components $D_z$ and $B_z$ that stay constant throughout a thin layer, the normal force can be expressed as
\begin{multline}\label{Finterface}
{\bar{F}}_{\rm int}=\int_{-\delta/2}^{\delta/2} \bar{F}_z dz=\frac{1}{8\pi}[(\varepsilon_1-\varepsilon_2)\overline{E_\tau^2}+ (\mu_1-\mu_2)\overline{H_\tau^2}+\\
\left(1/\varepsilon_2- 1/\varepsilon_1\right)\overline{D_z^2}+\left(1/\mu_2-1/\mu_1\right)\overline{B_z^2}].
\end{multline}
It does not depend on the particular structure of the transient layer and can be evaluated, for example, for the fields arising  when an arbitrarily polarized EM-wave is incident on the interface at an arbitrary angle.
Simple calculations show that this force coincides with that evaluated from  the simple 'corpuscular' relation
\begin{equation}\label{Finterface_kinematic}
{\bf \bar F}=\hbar{\bf k}_{0}N_{0z}-\hbar{\bf k}_{t} N_{t z}-\hbar{\bf k}_{r}N_{r z},
\end{equation}
where ${\bf k}_{0,r,t}$ are the wavevectors of the incident, reflected and transmitted waves respectively, and ${ N}_{0,r,t\ z}$ are the $z$-components of the corresponding polariton number fluxes.

In a weakly lossy homogeneous medium, i.e., for $\mu=\mu'+i\mu''$ and $\varepsilon=\varepsilon'+i\varepsilon''$ with small imaginary parts, the complex representations of the oscillating fields: ${\bf E}=({\bf E}_0e^{-i\omega t}+ c.c.)/2$, ${\bf D}=(\varepsilon{\bf E}_0e^{-i\omega t}+ c.c.)/2$, etc.,
substituted in Eq.\eqref{Ftot1} yield after time-averaging
\begin{equation}\label{Floss}
\bar{\bf F}=\frac{\omega}{4\pi c}(\varepsilon'\mu''+\mu '\varepsilon'')\overline{(\bf E\times H)}.
\end{equation}
This force is always collinear with the average momentum density \eqref{G} and corresponds to the absorbed polariton momentum. Thus for a linearly polarized monochromatic plane wave it is proportional to the absorbed energy $Q=\omega(4\pi)^{-1}(\varepsilon''\overline{E^2}+\mu''\overline{H^2})$ as $\bar{\bf F}= Q {\bf k}/\omega$, which means that the absorption of energy quant $\hbar\omega$ is accompanied by the absorption of momentum quant $\hbar \bf k$.

For a finite slowly modulated pulse of monochromatic light propagating through isotropic non-lossy medium, one introduces complex slow varying amplitudes, e.g., the electric field
${\bf E}= 1/2[{\bf E}_0(t) e^{-i\omega t}+c.c.]$. The frequency dispersion modifies the relations between field amplitudes and inductions \cite{Landau} so that the slow-varying electric induction amplitude reads ${\bf D}_0(t)=\varepsilon{\bf E}_0(t)+i\ {d\varepsilon}/{d\omega}\ \dot{\bf E}_0(t)$ and
the slow amplitude of its first time derivative equals $\left({\partial\bf D}/{\partial t}\right)_0(t)=-i\omega\varepsilon{\bf E}_0(t)+{d(\omega\varepsilon)}/{d\omega}\ \dot{\bf E}_0(t)$.
Substituting these amplitudes together with the similar magnetic ones into Eq.\eqref{Ftot} yields a very simple result:
\begin{equation}
\bar{\bf F}=\frac{\partial \bar{\bf G}}{\partial t},
\end{equation}
where the time averaged  momentum density $\bar{\bf G}$ is given by Eq.~\eqref{G} taken with the slow varying field amplitudes. This represents a generalization of the Abraham force.

Similarly, one can evaluate the momentum transfer during light-light interactions in nonlinear media. For simplicity, consider interaction of collinear waves propagating along $z$-axis and polarized linearly along $x$-axis and assume that the energy exchange is dominated by a single channel for which the set of frequencies $\{\omega_n\}$ fulfills the condition $\sum_n l_n \omega_n=0$  with integer $l_n$ and the wavevectors are weakly mismatched: $\sum_n l_n k_n=q$. For definiteness, consider the dielectric nonlinearity with nonlinear dielectric susceptibility $\chi^{(M)}$ of the order $M$.

Then the electric field can be presented via the slow-varying with $z$ amplitudes: $E=\sum_n E_n=1/2\sum_n[E_{n0}(z)e^{ik_nz-i\omega_nt}+c.c.]$ which obey the system of equations:
\begin{equation}\label{Eslow}
 \frac{dE_{n0}}{dz}=\frac{2\pi i}{k_n}\frac{\omega_n^2}{c^2}\mu(\omega_n)P^{\rm NL}_{n0}.
\end{equation}
Here the slow-varying amplitude of nonlinear polarization at frequency $\omega_n$ reads $P^{\rm NL}_{n0}=2^{-M}e^{-{\rm sgn}(l_n) iqz}\chi_n^{(M)}(E_{n0})^{1-|l_n|}\prod_{n'\ne n} (E_{n'0})^{-{\rm sgn}(l_n) l_{n'}}$, the order of the nonlinearity is
$M=\sum_{n}\left|l_{n}\right|-1$, for negative field powers the complex conjugate amplitudes $E_{n0}^*$ must be substituted, and $\chi_n^{(M)}$ is the $x$-component of the susceptibility tensor taken at appropriate frequencies: $\chi_n^{(M)}=\chi_{x ...x}^{(M)}(\omega_n; \{\omega_{n'}\})$.

Upon substituting $D=\sum_n\varepsilon(\omega_n)E_n+4\pi P^{\rm NL}$ and $B=\sum_n\mu(\omega_n)H_n$ into the force \eqref{Ftot1} only the terms
\begin{equation}
\frac{1}{2} {\left(E\frac {dP^{\rm NL}} {dz} -P^{\rm NL}\frac {dE} {dz}\right)}
\end{equation}
contribute to the time average. Neglecting higher powers of nonlinearity one can conveniently express the force density by the polariton number density fluxes, $N_n=S_n/(\hbar\omega)=c/(8\pi\hbar\omega_n)\sqrt{\varepsilon(\omega_n)/\mu(\omega_n)}|E_{n0}|^2$:
\begin{equation}\label{fnl}
F_z=\sum_n\hbar k_n\frac{dN_n}{dz}-\frac{\hbar q}{2}\sum_n {\rm sgn}(l_n)\frac{dN_n}{dz}.
\end{equation}

For the perfect phase matching, $q=0$, only the first term is nonzero and the force has intuitively clear 'corpuscular' form compensating the possible momentum unbalance during conversion  and vanishing, for example, for phase-matched $M$-th harmonic generation. In presence of phase mismatch, however, the second term in Eq.~\eqref{fnl} gives a nonzero contribution as the medium acquires the mismatched momentum. This establishes the link between the phase matching and the momentum conservation, which has been intuitively adopted e.g. in Refs.~\onlinecite{Armstrong,Landau, Bahabad} and numerous others.

Finally, consider the implications of the above for LHM. First, the momentum density \eqref{G} is parallel to the Pointing vector in ordinary media and antiparallel in LHM ($\varepsilon<0$, $\mu<0$). Indeed, the involved frequency derivatives must stay positive to ensure the stability of the EM energy \cite{Landau}, and, therefore, the sign of the square brackets follows that of $\varepsilon$ and $\mu$. Next, the force acting on an interface between media of different handedness according to \eqref{Finterface} always points in the direction of the LHM. For a slab of lossless LHM, however, the forces from the input and output interfaces partially compensate to ensure the total momentum conservation. In a lossy LHM the force \eqref{Floss} is opposite to the Poynting vector. For phase matched harmonic generation in non-dissipative LHM, no pressure is exerted in the bulk although an actual reversal of the energy flux can take place if the medium is left-handed not for all waves. In the latter case for the material acting as a nonlinear mirror  \cite{SHGLHM} the total light pressure arises solely at the surfaces in accordance with Eq.~\eqref{Finterface_kinematic} at each frequency.

In conclusion, we have shown that although there exists a continuum of possibilities to describe the momentum exchange between EM radiation and continuous medium, one can formulate an approach that provides a simple and intuitively clear picture. It is compatible with the polariton momentum $\hbar \bf k$, yields the compact general form \eqref{G} of the momentum density of EM wave and allows simple representation of light pressure during reflection/refraction, absorption and nonlinear conversion. This enables one, in particular, to divide the total momentum transfer in a complex real situation into elementary processes yielding detailed understanding of the roles of key parameters and conditions.

We are grateful to our colleagues V.E. Dmitrienko, B.I. Sturman, and V.A. Chizhikov for stimulating discussions, criticism and helpful advices.

\end{document}